\documentclass[aps,preprint,showpacs,groupedaddress,floatfix]{revtex4}
\usepackage{graphicx}
\usepackage{dcolumn}
\usepackage{bm}

\begin{document}

\title{Spin-Isospin Resonances and Neutron Skin of Nuclei}
\author{D. Vretenar}
\affiliation{Physics Department, Faculty of Science, University of Zagreb, Croatia}
\author{N. Paar}
\affiliation{Physik-Department der Technischen Universit\"at M\"unchen, D-85748 Garching,
Germany and\\
Institut f\" ur Kernphysik, Technische Universit\" at Darmstadt Schlossgartenstr. 9,
64289 Darmstadt, Germany}
\author{T. Nik\v si\' c}
\affiliation{Physics Department, Faculty of Science, University of Zagreb, Croatia, and \\
Physik-Department der Technischen Universit\"at M\"unchen, D-85748 Garching,
Germany}
\author{P. Ring}
\affiliation{Physik-Department der Technischen Universit\"at M\"unchen, D-85748 Garching,
Germany}
\date{\today}

\begin{abstract}
The Gamow-Teller resonances (GTR) and isobaric analog states (IAS) 
of a sequence
of even-even Sn target nuclei are calculated by using the 
framework of the relativistic
Hartree-Bogoliubov model plus proton-neutron quasiparticle random-phase
approximation. The calculation reproduces the experimental data on 
ground-state properties, as well as the excitation energies of 
the isovector excitations. It is shown that the isotopic dependence of 
the energy spacings between the GTR and IAS provides direct information 
on the evolution of neutron skin-thickness along the Sn 
isotopic chain. A new method is suggested for determining the difference 
between the radii of the neutron and proton density distributions along 
an isotopic chain, based on measurement of the excitation energies 
of the GTR relative to the IAS.

\end{abstract}

\maketitle
The determination of neutron density distributions in nuclei
provides not only basic
nuclear structure information, but it also places important
additional constraints on effective interactions used in nuclear
models. Various experimental methods have been used, or suggested, 
for the determination of the neutron density and
differences between radii of the neutron and proton distributions 
\cite{Bat.89,Kra.91,Suz.95,Kra.99},
but no existing measurement of neutron densities or radii has an 
established accuracy of one percent~\cite{Hor.99}.
 
Potentially, a very accurate and model independent 
experimental method for the
determination of neutron densities is the elastic scattering of longitudinally
polarized electrons on nuclei. The electron interacts with a nucleus
by exchanging either a photon or a $Z^0$ boson. Parity violation arises 
from the interference of electromagnetic and weak neutral amplitudes.
The parity-violating
asymmetry parameter, defined as the difference between cross sections
for the scattering of right- and left-handed longitudinally
polarized electrons, provides direct information on the
Fourier transform of the neutron density~\cite{DDS.89}.
A recent extensive analysis of possible parity-violating
measurements of neutron densities and their theoretical interpretation 
can be found in Refs.~\cite{Hor.98,Hor.99,Vre.00}.
Neutron density 
measurements will be also crucial for the analysis of atomic 
parity-violating (APV) experiments~\cite{Hor.99,Mus.99,PW.99,VLR.00}. 
Measurements of parity non conservation  effects in
intermediate and heavy atomic systems might provide very stringent
tests of the Standard Model of electroweak interactions.
At the present level of APV experimental precision, the dominant inaccuracy 
in the interpretation of the data is associated with atomic theory 
uncertainties in the electron density at the nucleus.
These uncertainties could be considerably reduced by measuring 
APV observables for different atoms along an isotopic chain. Experiments
involving isotope ratios, however, will be very sensitive to changes
in the neutron density distribution along the isotopic chain. 
For atomic PV experiments it would be 
very important to determine with high accuracy ($\approx 1$\%) 
the neutron spatial distribution for a sequence of isotopes~\cite{VLR.00}.
 
Krasznahorkay {\em et al.} have used the excitation of the giant dipole
resonance (GDR) \cite{Kra.91} and the excitation of the 
spin-dipole resonance (SDR) \cite{Kra.99} to extract the neutron-skin 
thickness of nuclei. In Ref.~\cite{Kra.99}, in particular, it has been 
demonstrated that there is a predictable correlation between the SDR 
cross section and the difference
between the rms radii of the neutron and proton density distributions.
By normalizing the results in the case of $^{120}$Sn, data on neutron-skin
thickness along the stable Sn isotopic chain were obtained, in good
agreement with theoretical predictions.

In this Letter we suggest a new method for determining the difference 
between the radii of the neutron and proton density distributions along 
an isotopic chain, 
based on measurement of the excitation energies of the Gamow-Teller
resonances relative to the isobaric analog states.  

Nucleons with spin up and spin down can oscillate either in 
phase (spin scalar S=0 mode) or out of phase (spin vector S=1 
mode). The spin vector, or spin-flip excitations can be of 
isoscalar (S=1, T=0) or isovector (S=1, T=1) nature. These 
collective modes provide direct information on the spin and
spin-isospin dependence of the effective nuclear interaction
(for an extensive review see Ref.~\cite{Ost.92}).
Especially interesting is the collective spin-isospin oscillation 
with the excess neutrons coherently changing the direction of their 
spins and isospins without changing their orbital motion -- the 
Gamow-Teller resonance (GTR) $J^\pi = 1^{+}$. The simplest 
charge-exchange excitation mode, however, does not require 
the spin-flip (i.e. S=0) and corresponds to the 
well known isobaric analog state (IAS) $J^\pi = 0^{+}$. 
The spin-isospin characteristics of the GTR and the IAS are
related through the Wigner supermultiplet scheme. The Wigner 
SU(4) symmetry implies the degeneracy of the GTR and IAS, and 
furthermore the resonances would completely exhaust the corresponding
sum rules \cite{Gap.74}. The Wigner SU(4) symmetry is, however, 
broken by the spin-orbit term of the effective nuclear potential. 
In Ref.~\cite{Gap.74} it was noted that the energy difference 
between the GTR and the IAS decreases with increasing asymmetry
$(N-Z)/A$.

It is implicit, therefore, that the energy difference between the 
GTR and the IAS reflects the magnitude of the effective spin-orbit
potential. In the framework of relativistic mean-field (RMF) theory
the spin-orbit term of the effective single-nucleon potential
displays a strong isospin dependence: the magnitude of the 
spin-orbit potential is considerably reduced in neutron-rich  
nuclei\cite{LVR.97}. This results
in a reduction of the energy splittings between spin-orbit partner states.
In Ref.~\cite{LVR.98} the relativistic 
Hartree-Bogoliubov (RHB) model was applied in
the description of ground-state properties of Ni and Sn isotopes.
The NL3 parameter set \cite{LKR.97} was used for the effective 
mean-field Lagrangian, and pairing correlations were described
by the pairing part of the finite range 
Gogny interaction D1S~\cite{BGG.84}. Fully self-consistent
RHB solutions were calculated for the Ni ($28\leq N\leq 50$) 
and Sn ($50\leq N\leq 82$) isotopes. The resulting binding energies, 
neutron separation energies, and proton and neutron rms radii
were found in excellent agreement with available
experimental data. With the increase of the number of neutrons,
the theory predicts a strong 
reduction of the spin-orbit potential (up to  $\approx 30$\% 
in the surface region). The location of the minimum of 
the spin-orbit potential is also shifted outwards, and this is 
reflected in the larger spatial extension of the scalar and vector densities,
which become very diffuse on the surface. The neutron-skin increases 
correspondingly. 
The issue of isospin dependence of the spin-orbit potential is, however,
not quite settled among the various theoretical
approaches (relativistic and non-relativistic) \cite{NP.98}, 
and it is at the center of active experimental investigations in 
unstable nuclei.

In the following we shall demonstrate that there is a 
direct connection between the increase of the neutron-skin
thickness in neutron-rich nuclei, and the decrease of
the energy difference between the GTR and the IAS. The  
calculation is performed in the framework of the fully 
self-consistent RHB plus proton-neutron relativistic QRPA model. The
RHB model represents a relativistic extension of the Hartree-Fock-Bogoliubov
framework, and it provides a unified description of particle-hole 
($ph$) and particle-particle ($pp$) correlations, that is essential
for a quantitative analysis 
of ground-state properties and multipole response of unstable,
weakly-bound nuclei far from the line of $\beta$-stability. Another
relativistic model, the relativistic random phase approximation (RRPA), has
been recently employed in quantitative analyses of 
collective excitations in nuclei. 
In Ref.~\cite{Paa.03} we have formulated 
the relativistic quasiparticle random phase approximation (RQRPA) 
in the canonical single-nucleon basis of the relativistic
Hartree-Bogoliubov (RHB) model. By definition, the
canonical basis diagonalizes the density matrix and it is always localized. 
This particular representation of the RQRPA is
very convenient because, in order to describe transitions to
low-lying excited states in weakly-bound nuclei, the
two-quasiparticle configuration space must include states with both nucleons
in the discrete bound levels, states with one nucleon in a bound level and
one nucleon in the continuum, and also states with both nucleons in the
continuum. The relativistic QRPA of Ref.~\cite{Paa.03} is fully self-consistent.
For the interaction in the particle-hole channel effective Lagrangians with
nonlinear meson self-interactions are used, and pairing correlations are
described by the pairing part of the finite range Gogny interaction. Both in
the $ph$ and $pp$ channels, the same interactions are used in the RHB
equations that determine the canonical quasiparticle basis, and in the
matrix equations of the RQRPA. The RQRPA configuration space 
includes also the Dirac sea of negative energy states.

In the meson-exchange picture of nucleon-nucleon forces it is 
the $\pi$- and $\rho$-meson exchange that generate the spin-isospin 
dependent interaction terms. Even though in the RMF
description of the nuclear ground state the direct one-pion 
contribution vanishes at the Hartree level because of parity conservation, 
the pion nevertheless plays an important role for excitations 
that involve spin degrees of freedom. Since it has a relatively small mass, 
the pion mediates the effective nuclear interaction over large distances. 
In the present relativistic description of spin-flip and isospin-flip 
excitations we employ the $\pi$- and $\rho$-meson exchange in 
the $p-h$ residual interaction of the proton-neutron RQRPA. 
The first relativistic RPA calculations of isobaric analog resonances 
and Gamow-Teller resonances have been performed only recently \cite{Con.98}. 
This analysis was, however, 
restricted to doubly closed-shell nuclei. A rather small 
configuration space was used and, furthermore, configurations that 
include empty states from the negative-energy Dirac sea were 
neglected. A more complete relativistic RPA calculation of  
GT resonances in doubly closed-shell nuclei was reported in 
Ref.~\cite{Ma.03}. The ground states of 
$^{48}$Ca, $^{90}$Zr and $^{208}$Pb were calculated in the 
relativistic mean-field (RMF) model with the NL3 effective interaction.
The spin-isospin correlations in the RRPA calculation of the GT 
response functions were induced by the isovector mesons $\pi$ and $\rho$.
In addition to the standard non-linear NL3 effective interaction with the 
vector rho-nucleon coupling,
the effective Lagrangian included the pseudo-vector pion-nucleon 
interaction, but not the rho-nucleon tensor term. Although the 
pion does not contribute in the RMF Hartree calculation of the ground state,
it has a pronounced effect on the spin-isospin excitations. However, 
because of the derivative type of the pion-nucleon coupling, it is 
also necessary to include a zero-range Landau-Migdal term that
accounts for the contact part of the nucleon-nucleon interaction.  
The analysis of Ref.~\cite{Ma.03} has shown that 
the RRPA calculation with the NL3 
effective interaction, the pseudo-vector pion-nucleon coupling 
($m_{\pi}=138$ MeV and $f_{\pi}^{2}/{4\pi}=0.08$), and 
the Landau-Migdal force with the strength parameter $g_{0}^{\prime}= 0.6$,
nicely reproduces the experimental excitation energies of 
the main components of the GT resonances in $^{48}$Ca, $^{90}$Zr 
and $^{208}$Pb.  

In the present analysis we extend the approach of Ref.~\cite{Ma.03} and
employ the self-consistent RHB plus proton-neutron relativistic QRPA to 
calculate the GTR and IAS in the Sn isotopic chain. The RMF effective 
interaction is NL3, the pion-nucleon interaction Lagrangian reads
\begin{equation}
\mathcal{L}_{\pi N}=-\frac{f_{\pi}}{m_{\pi}}\bar{\psi}\gamma_{5}\gamma_{\mu
}\partial^{\mu}{\vec{\pi}}{\vec{\tau}}\psi \; ,
\end{equation}
and we also include the Landau-Migdal zero-range force in the 
spin-isospin channel of the residual $p-h$ interaction:
\begin{equation}
V(1,2)=g_{0}^{\prime}\left(  \frac{f_{\pi}}{m_{\pi}}\right)  ^{2}\vec{\tau
}_{1}\cdot\vec{\tau}_{2}~\mathbf{\Sigma}_{1}\cdot\mathbf{\Sigma}_{2}%
~\delta(\bm r_{1}- \bm r_{2}) \; .
\label{deltapi}%
\end{equation}
With $g_{0}^{\prime}= 0.6$ our model calculation 
reproduce the results of Ref.~\cite{Ma.03} for the GTR in doubly 
closed-shell nuclei. For open-shell nuclei 
pairing correlations are described by the pairing part of the 
finite range Gogny interaction D1S. As it has been shown in 
the RHB+RQRPA analysis of Ref.~\cite{Paa.03}, the consistent 
inclusion of the pairing interaction both in the static RHB and in 
the dynamical linear response, is essential in order to 
satisfy the energy weighted sum rules and for the decoupling of 
spurious modes. In the present calculation we neglect
possible proton-neutron pairing correlations. Even though 
these correlations can contribute to the low-lying GT strength responsible 
for $\beta$-decay, their effect on the main component of the GTR 
can be safely neglected.  

In Fig.~\ref{figA} we display the calculated differences between 
the excitation energies of the main component of the Gamow-Teller 
resonances and the respective isobaric analog states for the
sequence of even-even Sn target nuclei with $A=112 - 124$. The 
result of fully self-consistent RHB plus proton-neutron RQRPA 
calculations are shown in comparison with experimental data 
obtained in a systematic study of the ($^3$He,t) charge-exchange 
reaction over the entire range of stable Sn isotopes \cite{Pham.95}. 
The calculated energy spacings are in very good agreement with 
the experimental values, and this result clearly demonstrates that 
the relativistic mean-field plus RPA framework provides a very 
natural description of nuclear spin and isospin excitations.
For lighter Sn isotopes it appears that the calculated values differ
somewhat from the experimental trend but, of course, the theoretical 
energy spacings might depend on the details of the effective
interaction. In fact, the experimental data shown in Fig.~\ref{figA}
provide very valuable information that can be used in constraining
the spin-isospin channel of the effective interaction. In order 
to illustrate the agreement between the data and model results,
in the insert of Fig.~\ref{figA} the calculated excitation energies of the 
isobaric analog states are compared with experiment \cite{Pham.95}.
Except for $^{112}$Sn, the agreement between data and the calculated
values of $E_{\rm IAR}$ is better than $\approx 0.5$ MeV.

In Fig.~\ref{figB} the calculated and experimental 
energy spacings between the 
GTR and IAS are plotted as a function of the calculated 
differences between the rms radii of the neutron and proton 
density distributions of even-even Sn isotopes (upper panel). 
The calculated radii correspond to the RHB NL3+D1S self-consistent 
ground-state solutions, on
which the proton-neutron RQRPA calculations are performed.
In the lower panel the calculated differences between neutron and  
proton rms radii are compared with available experimental data 
\cite{Kra.99}. In Fig.~\ref{figB}
we notice a remarkable uniform dependence of the energy 
spacings between the GTR and IAS on the size of the neutron-skin.
In principle, therefore, the value of $r_n - r_p$ can be directly determined
from the theoretical curve for a given value of 
$E_{\rm GT} - E_{\rm IAS}$. This method is, of course, not
completely model
independent, but it does not require additional assumptions. 
Since the neutron-skin thickness is determined in an indirect 
way from the measurement of the GTR and IAS excitation 
energies in a sequence of isotopes, in practical applications
at least one point on the theoretical curve should be checked 
against independent data on $r_n - r_p$.

In conclusion, we have employed the self-consistent 
RHB plus proton-neutron relativistic QRPA to calculate the 
spin-isospin excitations in a sequence of Sn isotopes. By using 
the NL3 parameter set for the effective RMF Lagrangian, 
the pairing part of the Gogny interaction D1S for the $T=1$ pairing 
channel, the pseudo-vector pion-nucleon 
interaction and the Landau-Migdal zero-range force with
$g_{0}^{\prime}= 0.6$ in the
spin-isospin channel of the residual
$p-h$ interaction of the proton-neutron RQRPA,
it has been possible to reproduce experimental data on 
the excitation energies of the Gamow-Teller
resonances relative to the isobaric analog states. We have 
also shown that the isotopic dependence of the energy 
spacings between the GTR and IAS provides direct information 
on the evolution of neutron skin-thickness along the Sn 
isotopic chain. Very good results have been obtained in comparison
with available data on $r_n - r_p$. This analysis suggests that the
neutron-skin thickness can be determined from the measurement 
of the excitation energies of the GTR relative to IAS.
\bigskip \bigskip

\newpage

\begin{figure}
\caption{ RHB plus proton-neutron RQRPA results for the energy spacings between 
the excitation energies of the main component of the Gamow-Teller 
resonances and the respective isobaric analog states for the 
sequence of even-even $^{112 - 124}$Sn target nuclei. 
The experimental data are from Ref.~\protect\cite{Pham.95}.
In the insert the calculated excitation energies of the 
isobaric analog states are compared with data.}
\label{figA}
\end{figure}

\begin{figure}
\caption{The proton-neutron RQRPA and experimental \protect\cite{Pham.95}
differences between the excitation energies
of the GTR and IAS as a function of the calculated 
differences between the rms radii of the neutron and proton 
density distributions of even-even Sn isotopes. (upper panel) 
In the lower panel the calculated differences $r_n - r_p$
are compared with experimental data \protect\cite{Kra.99}.}
\label{figB}
\end{figure}

\end{document}